# Epitaxial growth of atomically thin $Ga_2Se_2$ films on c-plane sapphire substrates

Mingyu Yu,[1] Lottie Murray,[1] Matthew Doty,[1] and Stephanie Law [1,2,3a)]

[1]Department of Materials Science and Engineering, University of Delaware, 201 Dupont Hall, 127 The Green, Newark, Delaware 19716
[2]Department of Physics and Astronomy, University of Delaware, 217 Sharp Lab, 104 The Green, Newark, Delaware 19716
[3]Department of Materials Science and Engineering, Pennsylvania State University, N-232 Millennium Science Complex, University Park, PA 16802

a) Electronic mail: slaw@psu.edu

## ABSTRACT

Broadening the variety of two-dimensional (2D) materials and improving the synthesis of ultrathin films are crucial to the development of the semiconductor industry. As a state-of-the-art 2D material, $Ga_2Se_2$ has attractive optoelectronic properties when it reaches the atomically-thin regime. However, its van der Waals epitaxial growth, especially for the atomically-thin films, has seldom been studied. In this paper, we used molecular beam epitaxy to synthesize $Ga_2Se_2$ single-crystal films with a surface roughness down to 1.82 nm on c-plane sapphire substrates by optimizing substrate temperature, Se:Ga flux ratio, and growth rate. Then we used a 3-step mode to grow $Ga_2Se_2$ films with a thickness as low as 3 tetralayers and a surface roughness as low as 0.61 nm, far exceeding the performance of direct growth. Finally, we found that the surface morphology strongly depends on the Se:Ga flux ratio, and higher growth rates widened the suitable flux ratio window for growing $Ga_2Se_2$. Overall, this work advances the understanding of the vdW epitaxy growth mechanism for post-transition metal monochalcogenides on sapphire substrates.

## I. INTRODUCTION

Van der Waals (vdW) materials provide a promising platform for the fabrication of heterostructures capable of multiple functionalities.[1-5] The weak interlayer interactions in vdW crystals result in self-passivated surfaces with no dangling bonds which enable the stacking of different materials with different lattice constants with an ultimate thickness scaling down to the atomic level.[6,7] Following graphene,[6,8,9] a wide range of vdW materials such as topological insulators,[10-12] transition metal dichalcogenides (TMDs),[13,14] and post-transition metal chalcogenides,[1,15,16] were explored to expand the family of vdW materials.[6,17-19] The boom in vdW materials has triggered innovations in many fields such as photocatalysis,[18,20] optoelectronics,[19,21] and photovoltaics.[22] Among the vdW materials, $Ga_2Se_2$ deserves particular attention due to its exotic electronic and optoelectronic properties.

Compared to TMDs, $Ga_2Se_2$ has an anomalous bandgap behavior, i.e., it changes from a direct bandgap bulk material to an indirect bandgap monolayer.[1,23,24] The top of the valence band in the $Ga_2Se_2$ monolayer exhibits a "Mexican hat" dispersion,[24-26] resulting in a high density of states. Thus $Ga_2Se_2$ is an attractive two-dimensional (2D) material for the semiconductor industry.[27-30] Ga-Se compounds have two common phases: $Ga_2Se_2$ and $Ga_2Se_3$.[31] $Ga_2Se_2$ is the focus of this study because in the crystal structure of $Ga_2Se_3$, one third of the gallium sites are typically occupied by undesired vacancies.[32] A primitive $Ga_2Se_2$ cell consists of two hexagonal gallium planes sandwiched by two hexagonal selenium planes, called a tetralayer (TL), with a thickness of 8 Å.[33,34] There are four common polytypes of $Ga_2Se_2$ designated as β (2R)-, ε (2H)-, γ (3R)-, and δ (4H)-$Ga_2Se_2$, arising from the stacking sequence of AA'AA'..., ABAB..., ABCABC... and AA'B'BAA'B'B..., respectively.[35-38] ε- $Ga_2Se_2$ is the focus of this study since it is the most extensively explored polytype.

There are a variety of ways to synthesize vdW materials such as exfoliation from a bulk crystal [39-41] and chemical vapor deposition.[42,43] However, molecular beam epitaxy (MBE) synthesis has the advantages of extremely high sample purity, precise control of doping and layer thickness, *in-situ* monitoring of the growth surface, and a wafer-scale platform commonly used in industry.[31] Epitaxial growth of $Ga_2Se_2$ has been studied by the scientific community for decades.[31,44-48] The multiple phases and polytypes of the Ga-Se compounds make it challenging to achieve accurate $Ga_2Se_2$ stoichiometry.[49-51] Moreover, there has been relatively little effort on how to grow atomically-thin $Ga_2Se_2$ films on c-plane sapphire (c-sapphire) substrates. In 1998, Chegwidden et al [48] developed a high-temperature annealing treatment to remove their initial $Ga_2Se_3$ film and obtain polycrystalline $Ga_2Se_2$ on a subsequent deposition on the same substrate. In 2015, Chia-Hsin et al [44] obtained a single crystal $Ga_2Se_2$ epilayer on c-sapphire by MBE. In 2017, Lee et al [31] grew ε-$Ga_2Se_2$, and they found that increasing the Se:Ga flux ratio resulted in a $Ga_2Se_3$ phase and a spotty RHEED pattern. They obtained $Ga_2Se_2$ crystalline films with a thickness of 75 nm and an RMS roughness as low as 2 nm. However, they also observed particles on their films using AFM and speculated that they were Ga droplets.

Compared to the growth on 3D substrates, such as GaAs, the advantage of growing $Ga_2Se_2$ on c-sapphire is the negligible influence of lattice mismatch and thermal expansion coefficient mismatch,[52] leading to fewer misfits and threading dislocations. C-sapphire wafers are also inexpensive, readily available, and commonly used for the vdW epitaxy of other materials. However, vdW growth may suffer from more 0D defects (e.g., vacancies), due to easier re-evaporation, and more 1D defects (e.g., grain boundaries), due to less control over domain orientation.[53-56]

In this paper, we study the growth of atomically thin (about 3 TL) $Ga_2Se_2$ single crystal films on c-sapphire using MBE. The growth window was optimized by growing thick (about 32 nm) $Ga_2Se_2$ films. At this stage, the substrate temperature, Se:Ga flux ratio, and growth rate were varied to obtain films with the desired phase, lowest surface roughness, and highest crystallinity, as well as to understand how these parameters impact the quality of the resulting films. We achieved $Ga_2Se_2$ crystals with an RMS roughness as low as 1.82 nm using an optimal set of parameters, i.e., a Se:Ga flux ratio of 20, a growth rate of 0.06 Å/s, and a growth temperature of 425 °C. These parameters were then used to grow atomically thin films. We developed a 3-step growth mode to obtain coalesced 3 TL-thick $Ga_2Se_2$ films with significantly improved surface morphology, in which the RMS roughness was reduced to 0.61 nm. Finally, we will discuss the presence of 3D features on the films, how they are related to vdW epitaxy, and how to reduce or eliminate them.

## II. EXPERIMENT

All samples were grown on epi-ready c-plane sapphire substrates using a dedicated Veeco GENxplor MBE system in the University of Delaware Materials Growth Facility. A Knudsen cell for gallium and a valved cracker cell for selenium were used for all growths, maintaining a temperature of 900 °C for the selenium cracking zone while controlling the selenium flux via the valve. Prior to growth, substrates were degassed in a load lock chamber at 200 °C and then transferred to a growth chamber where they were heated to 650 °C to desorb any residual contaminants. The substrates were then cooled to the desired growth temperature. Immediately before growth, the selenium shutter was opened 2 min before the gallium shutter to create a selenium atmosphere.[31,48] Then the Ga-Se films were grown via co-deposition; details are described in Section III. For the cooling process, contrary to our group's experiences growing other selenides like $Bi_2Se_3$[57-59] and $In_2Se_3$,[60] the exposure of $Ga_2Se_2$ to the selenium flux should be stopped immediately at the end of growth. If it is not, more 3D features will appear, as shown in the AFM images in Fig. S1 in the supplementary material at [URL will be inserted by AIP Publishing]. After growth, the samples were removed from the MBE and immediately vacuum packed to await analysis. All substrate temperatures were measured by a non-contact thermocouple. The reported source fluxes are beam equivalent pressures measured at the beginning of each growth day using a beam flux monitor.

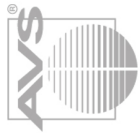

In this study, an *in-situ* reflection high-energy electron diffraction (RHEED) system was used to monitor the surface morphology and crystallinity during growth. Atomic force microscopy (AFM) and scanning electron microscopy (SEM) measurements were used to characterize the surface morphology. AFM was performed on a NeaSNOM Microscope at the Advanced Materials Characterization Laboratory and SEM was done on an Auriga 60 CrossBeam at the Keck Center for Advanced Microscopy and Microanalysis. X-ray reflectivity (XRR) was conducted on Rigaku Ultima IV XRD to determine the thickness of the deposited films. X-ray diffraction (XRD) 2θ/ω scans were performed on a Bruker D8 XRD using a Cu $K\alpha_1$ source to analyze the film phase, and ω scans were performed on a PANalytical 4-Circle X'Pert3 MRD to assess crystal quality. The absence of sapphire substrate peaks in our XRD patterns is ascribed to the rough alignment between the sample and the XRD system. Room temperature Hall effect measurements were conducted in the van der Pauw configuration to study the film electrical properties. Raman spectra were collected from a Horiba LabRAM HR Evolution for better identification of phases.

## III. RESULTS

Starting with a thick film allows for easier optimization of the following growth conditions: growth temperature, Se:Ga flux ratio (hereinafter referred to as flux ratio), and growth rate. At this stage, unless otherwise stated, all samples were grown using a gallium flux of approximately $2.6\times10^{-8}$ Torr equivalent to a 0.06 Å/s growth rate for 90 min, resulting in a thickness of 32 (±3) nm. Later, the optimal conditions were applied to grow ultrathin films approximately 3 TL thick.

### A. Se:Ga flux ratio

We first investigated the effect of flux ratio by fixing the growth temperature at 450 °C and the growth rate at 0.06 Å/s. Fig. 1 shows the XRD analysis of the samples grown using different flux ratios. Sample 1, grown using a ratio of 100, only presents a peak at 2θ=28.3° associated with the $Ga_2Se_3$ (111) orientation.[61] A ratio of 60 used for Sample 2 results in an amorphous phase. The two samples indicate that an excess of selenium leads to a $Ga_2Se_3$ or amorphous phase. Samples 3, 4, and 5 grown using the ratios of 37, 22 and 18, respectively, all show similar XRD curves, in which the two peaks at 2θ=11.1° and 22.2° correspond to the $Ga_2Se_2$ (002) and (004) orientations, respectively.[51] The three $Ga_2Se_2$ samples behaved significantly differently in the AFM scans, as shown in Fig. 2: Sample 3 consists of randomly distributed poorly coalesced domains, while Sample 5 has large clusters on the surface. An intermediate flux ratio of 22 (Sample 4) produced the most continuous film with an RMS roughness of 2.35 nm.

An unusual and unfortunate finding compared to our experience growing $Bi_2Se_3$[57-59] and $In_2Se_3$,[60] as well as others growing $Ga_2Se_2$,[31,45,70] is the extremely narrow flux ratio window within which smooth $Ga_2Se_2$ crystalline films can be obtained. For example, comparing Samples 4 and 5, although both flux ratios of 22 and 18 produced $Ga_2Se_2$ single crystals, the slightly smaller ratio led to large clusters as shown in Fig. 2(c). These clusters may be caused by gallium aggregation in the early stages of growth due to a slight selenium deficiency. More examples are shown in Fig. S2 and S3 in the supplementary material at [URL will be inserted by AIP Publishing]. Fig. S2 shows that when films are grown at 0.06 Å/s with a lower substrate temperature of 425 ºC, the appropriate flux ratio window is even narrower. Fig. S3 compares the surface morphology of samples grown using a ratio of 28 and 23, respectively, based on a higher growth rate of 0.11 Å/s. Sample 20 using a ratio of 23 displays a much smoother surface compared to Sample 19 using a ratio of 23, again confirming the stringent requirement for a suitable flux ratio. This extremely high sensitivity to flux ratio may be due to the multiple phases of Ga-Se compounds and the use of a selenium cracker cell. It is vital that the selenium flux is not in excess to stop the formation of $Ga_2Se_3$ phases, insufficient selenium will result in gallium agglomerates, especially on the c-sapphire substrate. Therefore, it is more difficult to achieve the desired stoichiometry of $Ga_2Se_2$ than to obtain $Bi_2Se_3$ or $In_2Se_3$, since both of those materials can be grown with a large selenium excess without the formation of unwanted phases. In addition, the cracked selenium molecules we used have a higher reactivity and incorporation efficiency than the uncracked ones, which may place more pressure on the precision of the Se:Ga flux ratio.

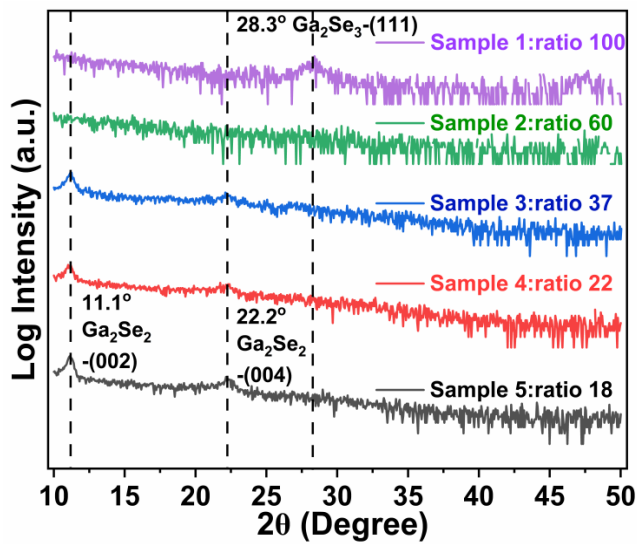

**FIG. 1.** XRD 2θ/ω scans of the samples grown using different Se:Ga flux ratios at a substrate temperature of 450 ºC and a growth rate of 0.06 Å/s.

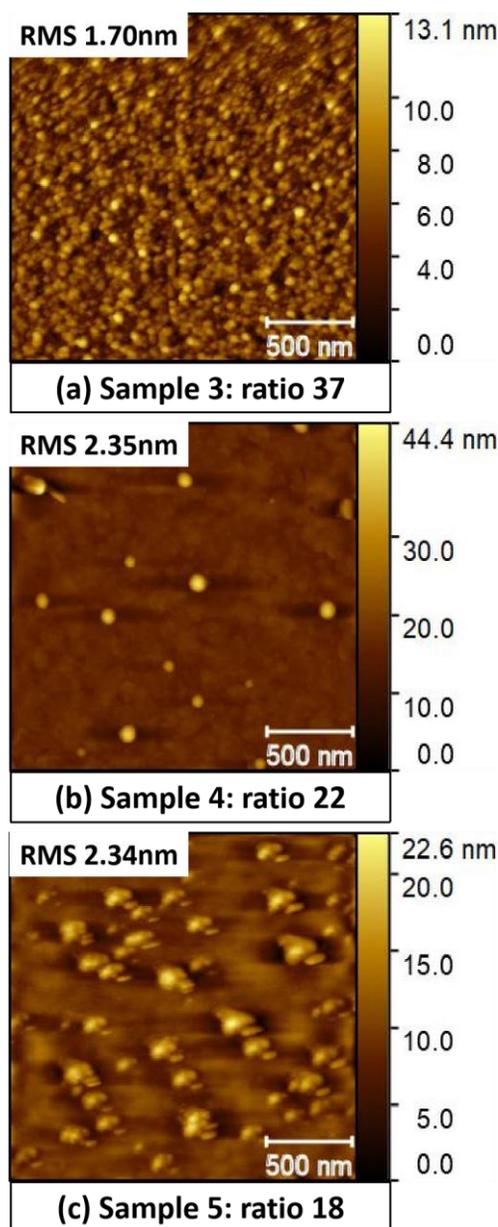

**FIG. 2.** AFM images of Sample 3 (a), Sample 4 (b) and Sample 5 (c), grown using a Se:Ga flux ratio of 37, 22, and 18, respectively, at a growth temperature of 450 °C. The scan size is 2×2 μm² for all samples. For Sample 5, multiple scans were taken and they all show similar surface morphology. Additional AFM images for Sample 5 are shown in Fig. S4 in supplementary material at [URL will be inserted by AIP Publishing].

### B. Substrate temperature

Next, we explored the effect of substrate temperature on film morphology. In this section, 37 was chosen as the flux ratio instead of 22 because Sample 4, despite showing a more continuous surface, has undesired 3D features. The main goal of this portion of the study was to obtain continuous films without 3D features by optimizing the growth temperature. In Fig. 3, neither Sample 6 grown at 480 °C, nor Sample 8 grown at 400°C show visible peaks in the XRD curves, indicating that a too high or a too low substrate temperature will lead to amorphous films. $Ga_2Se_2$ is known to sublimate around 500 °C as $Ga_2Se_{1/2}+Se_2$,[48] which sets a ceiling for the growth temperature. Furthermore, the hotter the substrate, the less likely the adatoms or compounds are to adhere to the growth surface, especially for vdW materials due to a weak interaction between the substrate and the deposited materials. However, low-temperature growth causes the problem of reduced adatom mobility,[57] resulting in more and smaller nucleation domains, which is a typical source of grain boundary defects. Fig. 3 shows that $Ga_2Se_2$ single crystal films can be obtained in the window of 425-450 °C. From the AFM images in Fig. 4, we see that Sample 7 grown at a relatively low temperature of 425°C has a more continuous surface than Sample 3 grown at 450 °C, although it has 3D features. By reducing the flux ratio to 20 at 425 °C, the density and height of 3D features were minimized as shown in Fig. 4(e). Details on optimizing the flux ratio at the growth temperature of 425 °C are described in Fig. S2.

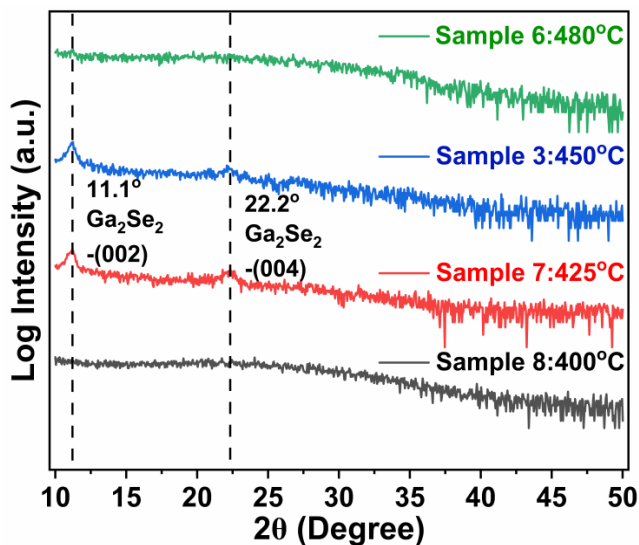

**FIG. 3.** XRD 2θ/ω scans of the samples grown at varying substrate temperatures using a flux ratio of 37 and a growth rate of 0.06 Å/s.

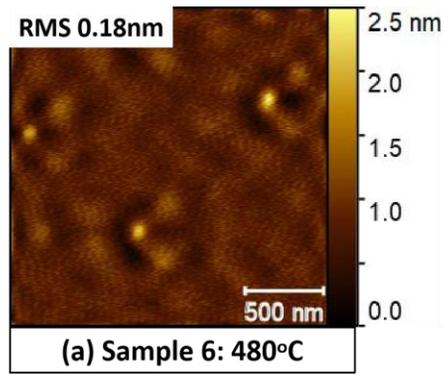
(a) Sample 6: 480°C, RMS 0.18nm

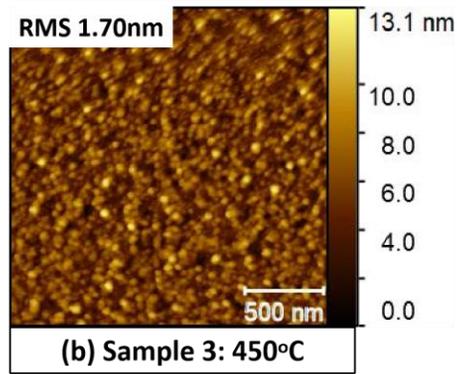
(b) Sample 3: 450°C, RMS 1.70nm

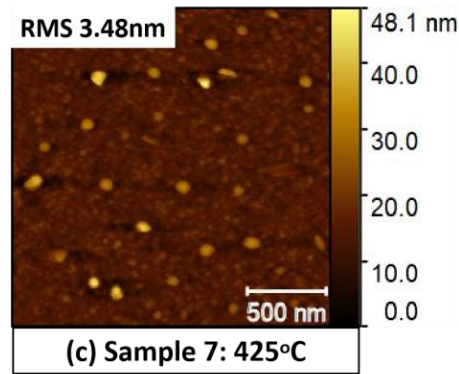
(c) Sample 7: 425°C, RMS 3.48nm

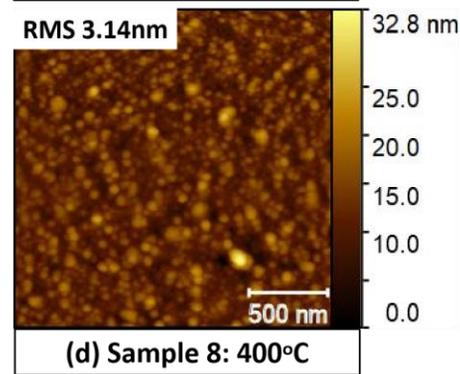
(d) Sample 8: 400°C, RMS 3.14nm

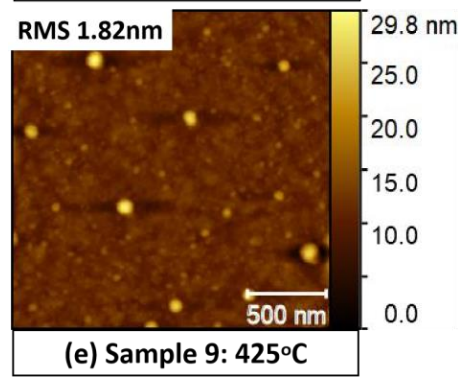
(e) Sample 9: 425°C, RMS 1.82nm

**FIG. 4.** AFM images of Sample 6 (a), Sample 3 [b, same as Fig. 2(a)], Sample 7 (c), and Sample 8 (d), grown at substrate temperatures of 480 ºC, 450 ºC, 425 ºC, and 400ºC, respectively, fixing the flux ratio at 37. AFM image of Sample 9 (e), using the same growth temperature as Sample 7 but with an optimized flux ratio of 20. The scan size is 2×2 μm$^2$ for all samples.

### C. Growth rate

The last parameter we studied to optimize the growth of thicker $Ga_2Se_2$ films was the growth rate. Theoretically, a slow growth rate is beneficial for the formation of single crystal films with fewer defects, as it allows more time for adatoms reaching the growth surface to diffuse and nucleate domains, thus yielding larger but fewer domains and suppressing grain boundary defects or clusters. The ω scans depicted in Fig. S5 (j) in supplementary material at [URL will be inserted by AIP Publishing] confirm that for the vdW growth of $Ga_2Se_2$, a slower growth rate leads to a higher crystal quality. Therefore, 0.06 Å/s, the lowest stable growth rate our MBE system can achieve, was chosen for this study.

Another interesting finding is that growth rate significantly affects the necessary Se:Ga flux ratio, which can be derived from the XRD curves in Fig. 5. A flux ratio as high as 60 accompanied by a high growth rate of 0.45 Å/s yields a $Ga_2Se_2$ phase, while maintaining that high flux ratio but reducing the growth rate to 0.06 Å/s results in an amorphous phase. To regain the desired $Ga_2Se_2$ phases, the corresponding flux ratio must be reduced, such as the ratio of 22 used for Sample 4. It can be deduced that higher growth rates broaden the window of flux ratios in which $Ga_2Se_2$ single crystals can be achieved. This is surprising, as typically the flux ratio required to grow a compound is not strongly dependent on the growth rate. For these materials, the likely reason is that the lower growth rate leaves more time for gallium adatoms to migrate across the growth surface and diffuse over longer lengths to nucleate with further selenium atoms, resulting in a smaller amount of selenium required.

Furthermore, Sample 21, which was grown under optimal conditions, was selected for Φ scans (Fig. S6 in supplementary material at [URL will be inserted by AIP Publishing]). The Φ scan of $Ga_2Se_2$ revealed a 6-fold pattern at about every 60º, consistent with the expected hexagonal crystal structure, but indicating the presence of twin domains. Although an epitaxial relationship between the $Ga_2Se_2$ layer and the sapphire substrate is observed, the film displays substantial in-plane rotational disorder, suggesting a weak binding between the $Ga_2Se_2$ layer and the sapphire. Finally, Samples 4 and 9, which exhibit relatively optimal surface morphology, were chosen for Hall effect measurements. Both samples were found to be insulating with infinite resistance, contradicting previous research[5,31,33] that found $Ga_2Se_2$ to be p-type. This discrepancy could be attributed to the defects within the samples or a poor connection between the indium contact and the film.

Photoluminescence (PL) measurements were taken with the goal of analyzing the material bandgap, but the results were inconclusive. The PL spectra for all samples were very broad, with a spectral full-width half-max (FWHM) of 115 nm that is blueshifted by 40 nm relative to the expected GaSe PL peak centered on 620 nm with a 12 nm full-width half-max [71]. Moreover, there is significant variation in PL intensity across all of the samples, which could be indicative of inhomogeneity in the number of defects, grain boundaries, or structure. We suspect that the blue shift of the center of the PL peak, the extreme broadening of the PL peak, and the inhomogeneity of the PL intensity across the sample may all be due to the existence of the 3D features, which may be causing changes in strain, plasmonic enhancement, oxidation, or thin film interference effects. Because the 3D features are much smaller than the diffraction limited spot size of the optical microscope, and are measured only by AFM, we cannot directly correlate any variations in PL to the proximity of specific 3D features. Further study will be needed to isolate and understand these effects. For the purposes of optimizing growth conditions, which is the focus here, we considered only the average PL intensity across each sample as an indicator of film quality.

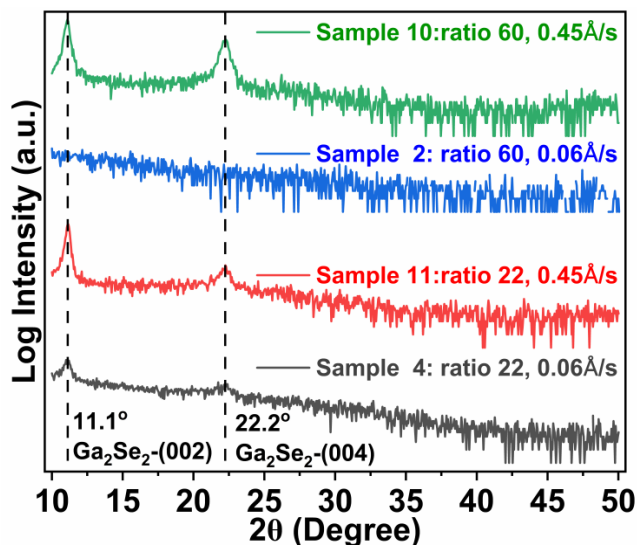

**FIG. 5.** XRD 2θ/ω scans of the Sample 10 (green, grown using a flux ratio of 60 and growth rate of 0.45 Å/s), Sample 2 (blue, grown using a flux ratio of 60 and growth rate 0.06 Å/s), Sample 11 (red, grown using a flux ratio of 22 and growth rate 0.45 Å/s), and Sample 4 (grey, grown using a flux ratio of 22 and growth rate 0.06 Å/s). All samples were grown at 450 °C and have a thickness of 32 ($\pm$3) nm.

### D. 3-step mode to grow 3 TL $Ga_2Se_2$ film

The set of optimal growth conditions identified by growing thick $Ga_2Se_2$ films is: a Se:Ga flux ratio of 20, a substrate temperature of 425 °C, and a growth rate of 0.06 Å/s. These parameters were next applied to grow $Ga_2Se_2$ films with a thickness of 3 tetralayers (TL) by shortening the growth time to 400 s. However, both the RHEED pattern [Fig. 6(a)] and the AFM scan [Fig. 6(b)] indicate that the resulting Sample 12 is only a collection of discrete islands rather than a coalesced layer. It is expected that the weak interaction between the substrate and the deposited materials makes it difficult to form a coalesced film with atomic-scale thickness,[59] and this morphology may be caused by the poor wettability of gallium atoms on sapphire.[62] The surface chemistry of sapphire does not support gallium adatoms wetting the surface, thus the first arriving gallium adatoms tend to ball up to form discrete islands rather than spread out and nucleate $Ga_2Se_2$ domains. This interpretation is supported by the observation of more continuous films with fewer visible islands with increasing thickness. This may be because the initially formed $Ga_xSe_y$ changes the surface chemistry, allowing the later formed $Ga_xSe_y$ to be uniformly distributed and gradually cover the initial islands.

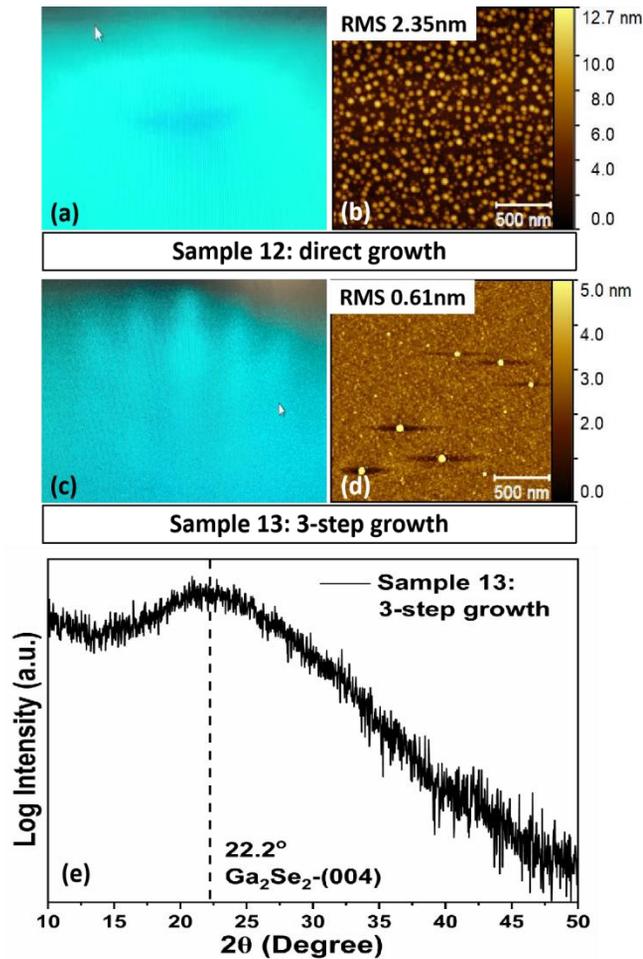

**FIG. 6.** RHEED (a) and AFM (b) images of Sample 12 grown directly using optimal conditions. RHEED (c), AFM (d), and XRD 2θ/ω (e) images of Sample 13 grown using the 3-step mode. AFM scan size is 2×2 μm$^2$ for all samples.

Based on these observations, we decided to pursue a 3-step growth mode, which was inspired from the MBE growth of other vdW films.[59,63] Fig. 7 illustrates the overall process, in which a growth rate of 0.06 Å/s and a flux ratio of 20 were adopted. In the first step, 20 TL $Ga_2Se_2$ were deposited on a sapphire substrate at 425 °C; then the substrate was heated to 1000 °C and annealed for 10 min in a selenium atmosphere to completely decompose the initial film. Decomposition was confirmed by the observation of only sapphire Kikuchi lines in the RHEED pattern. After that, the substrate was cooled to the desired growth temperature. Then we grew the first tetralayer of $Ga_2Se_2$ at a relatively low temperature of 300 °C. Afterwards, the substrate was heated to 450 °C to deposit the rest of the $Ga_2Se_2$ layers. In this way, at the end of the growth of 3 TL $Ga_2Se_2$, the RHEED shows a clear streaky pattern [Fig. 6(c)] associated with $Ga_2Se_2$ crystalline films,[1,32] which was in sharp contrast to the cloudy RHEED pattern [Fig. 6(a)] at the end of direct growth. This suggests a significant improvement of the resulting film quality, which is supported by the AFM image in Fig. 6(d). The morphology of the film grown using this 3-step mode is more typical of what is observed for thick $Ga_2Se_2$ films, and we can see a much-improved surface continuity with an RMS roughness as low as 0.61 nm. Notably, XRD scans [Fig. 6(e)] successfully detected a $Ga_2Se_2$ (004)-oriented peak from such a thin film. Although the peak of $Ga_2Se_2$ (002) orientation is not present, the peak at 2θ=22.2° confirms that this film is primarily $Ga_2Se_2$.

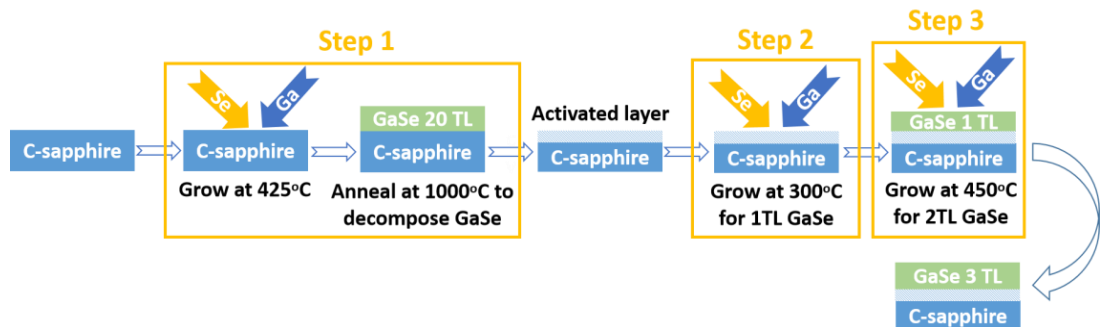

**FIG. 7.** A diagram of the process for 3-step mode to grow a 3 TL-Ga$_2$Se$_2$ film.

It is critical to understand the potential mechanism of this 3-step mode. Based on the experiences in similar systems (e.g., Ga$_2$Se$_2$ on silicon,[32] Ga$_2$Se$_2$ on sapphire,[48] Bi$_2$Se$_3$ on sapphire,[59] and WSe$_2$ on silicon[64]), it is suspected that the substrate pretreatment in Step 1 can alter the surface chemistry of sapphire to overcome the poor wettability of gallium atoms. Scott et al [48] used X-ray photoelectron spectroscopy to verify that a reacted interface layer containing traces of gallium and selenium was left on top after the evaporation of the initial film at high temperature. It is hypothesized that an exchange occurs between selenium and oxygen, as well as between gallium and aluminum. The weaker ionicity of selenium compared to oxygen results in a less polar, less reactive surface. Such passivated surfaces with fewer dangling bonds are favorable for vdW growth and may promote gallium atoms to wet the surface. Although the actual mechanism remains elusive, Fig. S7 in supplementary material at [URL will be inserted by AIP Publishing] demonstrates that the substrate-pretreatment (Step 1) plays a role in the growth of coalesced ultrathin Ga$_2$Se$_2$ films by comparing the AFM and RHEED images from Sample 23 (grown on a pristine substrate) and Sample 24 (grown on a pre-treated substrate). Because these films are otherwise identical, these data show that the pre-treatment step is critical. Next, a combination growth of low temperature (Step 2) followed by high temperature (Step 3) was inspired by the growth of Bi$_2$Se$_3$ on sapphire[57-59,65-67] since they both grow by vdW epitaxy. In general, the epitaxial growth of covalently bonded materials occurring at higher substrate temperatures results in larger but fewer domains and thus fewer defects, because the adatom mobility increases with substrate temperature. However, the story is opposite in vdW growth, in which higher substrate temperatures not only makes it more likely to re-evaporate the deposited film, but also causes gallium adatoms to diffuse faster and ball up with adjacent gallium adatoms, thereby yielding islands with lower substrate coverage. We believe that nucleating the film at a low temperature provides a platform for the further growth and coalescence of Ga$_2$Se$_2$ domains. Furthermore, the in-plane covalent bonds formed in the Ga-Se seed layer are stable up to the next high temperature step, resulting in a further improvement in surface morphology by increasing the adatom mobility.

## IV. DISCUSSION

One remaining question concerns the dot-like 3D features shown in the AFM images of Ga$_2$Se$_2$ samples, as exhibited in Fig. 2(b), 4(c)(e), 6(d), 8(a) and 9(b). Within the growth window where Ga$_2$Se$_2$ forms, no matter how we tuned the flux ratio, growth temperature, or growth rate, the 3D features could not be eliminated or significantly affected. The AFM and SEM images (Fig. S8 in supplementary material at [URL will be inserted by AIP Publishing]) only tell us that these 3D features are protrusions with spherical surfaces, ranging in diameter from 30 to 100 nm. The height of the 3D features on samples with a thickness around 32 nm ranges from 29 to 48 nm. A common perception is that a streaky RHEED pattern only appears when the surface of the grown film is smooth. However, taking Samples 4 and 9 shown in Fig. S9 in supplementary material at [URL will be inserted by AIP Publishing] as examples, we saw clear streaky RHEED patterns throughout the growth process, whereas the resulting film surface has many 3D features, which is contrary to our expectation. Based on these observations, two possible origins of the 3D features were proposed.

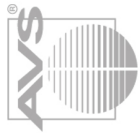

The first is that they may be impurities. If $Ga_2Se_2$ tends to adsorb impurities onto its surface, during the process of removing it from the MBE and performing AFM scanning, dust from the external environment may be absorbed and accumulated on the sample surface, which can explain why the RHEED pattern did not reflect those surface features. Based on this conjecture, we ultrasonically cleaned Sample 14 with N-methyl-2-pyrrolidone (NMP), a common solvent for removing organics. Fig. 8 compares the surface morphology before and after cleaning. There are visible features on the pristine sample as shown in Fig. 8(a). After sonication in NMP for 40 s, most of the 3D features disappeared, but ring-shape features were left in Fig. 8(b), and the RMS roughness was reduced to 1.14 nm. An amazing result was yielded by another 300 s of sonication. Fig. 8(c) shows no visible features and thus a smooth surface with an RMS roughness as low as 0.39 nm. In addition, we can see triangular $Ga_2Se_2$ nucleation domains. However, the SEM image in Fig. 8(d) demonstrates that a few 3D features persisted, and that this cleaning method left traces that appear as shadows in the SEM image and ring-shape features in the AFM image. While the 3D features can be ultrasonically removed to some extent, surface impurities should not be that difficult to remove. Moreover, the traces left after cleaning point imply that these features did not just adhere to the sample surface.

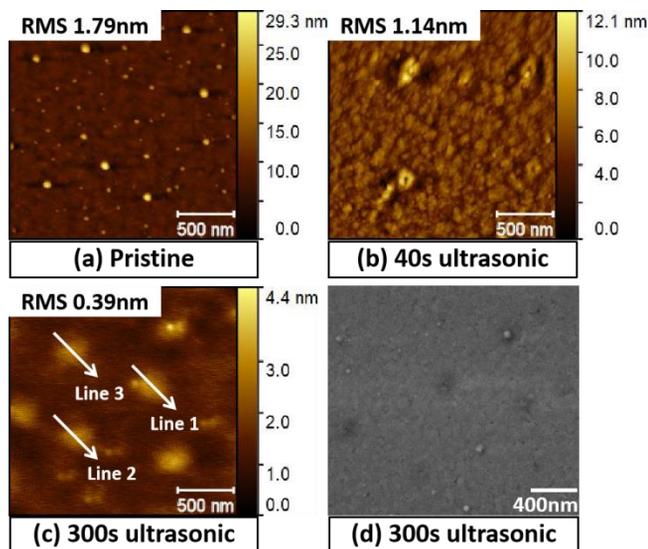

**FIG. 8.** AFM images of Sample 14 before cleaning (a), after ultrasonication in NMP for 40 s (b), and after ultrasonication in NMP for another 300 s (c). SEM image of Sample 14 after ultrasonication in NMP for another 300 s (340 s in total) (d). Sample 14 was grown at 425 °C, using a flux ratio of 20 and a growth rate of 0.12 Å/s, with a thickness of about 65 nm. AFM scan size is 2×2 μm$^2$ for all samples. Profile scans were taken along the three white arrows in (c), from which we know that the height differences along Lines 1, 2, and 3 are 7.1, 9.1 and 8.5 Å, respectively, all of which are close to the TL thickness of $Ga_2Se_2$, so we believe that these triangular features are $Ga_2Se_2$ domains. The profile scans are shown in Fig. S10 in supplementary material at [URL will be inserted by AIP Publishing].

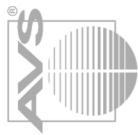

The evidence that 3D features remain after cleaning support our second conjecture, which is that the features formed at the bottom of the film, likely during the initial nucleation. To test this conjecture, we need to understand the composition of the features. One challenge is their small size: the largest feature is only about 0.1 μm in diameter. An investigation of the feature composition/phase was attempted on Sample 15, which, like other $Ga_2Se_2$ crystal samples, exhibited a streaky RHEED pattern during growth, pure XRD peaks of $Ga_2Se_2$ (002) and (004) orientations in Fig. 9(d), and the AFM image Fig. 9(b) with undesired 3D features. Raman tests were performed on a small area (1 μm diameter spot) and a large area (20×20 μm$^2$ square), respectively. Fig. 9(a) depicts the spectrum from a single spot in which the three peaks at 132 cm$^{-1}$, 210 cm$^{-1}$, and 310 cm$^{-1}$ correspond to the $A^1_{1g}$, $E^1_{2g}$, and $A^2_{1g}$ vibration modes of $Ga_2Se_2$, respectively.[1,31] The two peaks at 206 cm$^{-1}$ and 213 cm$^{-1}$ are thought to be a splitting of the 210 cm$^{-1}$ peak, which may be caused by surface strain or stress. The final two peaks at 155 cm$^{-1}$ and 249 cm$^{-1}$ can be attributed to $Ga_2Se_3$.[67,68] However, in a 20×20 μm$^2$ scan area, Fig. 9(c) shows that the three characteristic peaks of $Ga_2Se_2$ remain significant, but $Ga_2Se_3$ peaks are almost invisible (only a weak signal at 155 cm$^{-1}$ can be seen). The relatively weak intensity at 155 cm$^{-1}$ compared to 132 cm$^{-1}$ in Fig. 9(a), as well as the diminishing of $Ga_2Se_3$ signal over a large area both suggest that the film consists of a majority of $Ga_2Se_2$ and a minority of $Ga_2Se_3$. The $Ga_2Se_3$ phase is therefore likely to be too small to be detected by XRD or large-area Raman scans. From these data, we conclude that the 3D features are most likely $Ga_2Se_3$ inclusions. The formation mechanism may be that gallium adatoms ball up on the sapphire surface to form bumps at the initial stage of growth, providing more surface area for selenium atoms to incorporate, thus making the stoichiometry in these small regions closer to $Ga_2Se_3$ rather than $Ga_2Se_2$. This also explains the appearance of streaky RHEED patterns with 3D features on the growth surface. We hypothesize that the selenium cracker may also exacerbate the problem of 3D features by supplying more reactive selenium which may more easily form $Ga_2Se_3$ as noted previously.

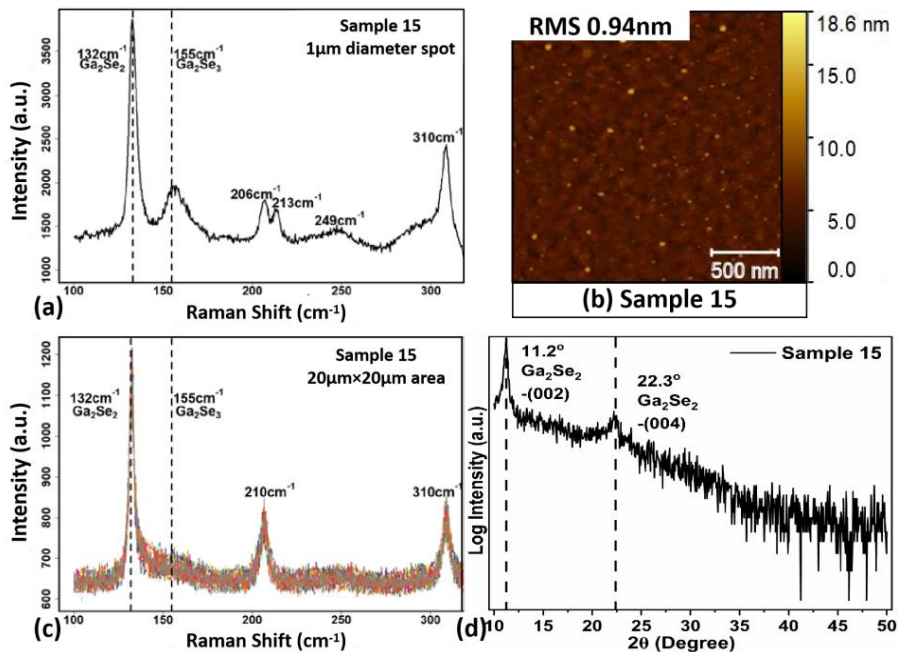

**FIG. 9.** Raman spectrum taken in a 1 μm diameter spot (a) and a 20×20 μm$^2$ area (c) of Sample 15, using a 532 nm laser, a 1800 groove/mm grating and a 100× objective. 100 mW and 5 mW laser powers were used for single spot scan and 20×20 μm$^2$ area scan,

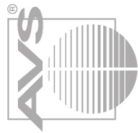

respectively. AFM image (b) and XRD 2θ/ω scan (d) of Sample 15. AFM scan size is 2×2 μm². Sample 15 was grown at 425 °C with a growth rate of 0.14 Å/s and a flux ratio of 20. The film thickness is about 77 nm.

## V. SUMMARY AND CONCLUSIONS

In this work, we studied the MBE growth of atomically-thin $Ga_2Se_2$ single-crystal films on c-plane sapphire. C-sapphire substrates are appealing due to their few dangling bonds, epi-ready availability, and relatively inexpensive prices. However, growth of $Ga_2Se_2$ on sapphire has rarely been studied. By examining thick films grown at varying temperatures, Se:Ga flux ratios, and growth rates, we found that the optimized growth window is: a substrate temperature of 425 °C, a Se:Ga flux ratio around 20, and a growth rate of 0.06 Å/s. The ceiling of the Se:Ga flux ratio at which $Ga_2Se_2$ single crystals can form increases with the growth rate, possibly because the slow growth allows gallium adatoms to diffuse farther to nucleate with selenium atoms, thereby reducing the necessary selenium flux. We also explored the growth of atomically-thin $Ga_2Se_2$ films. We found that direct growth of 3 TL $Ga_2Se_2$ using this window produced only uncoalesced films with many islands. By developing a 3-step growth mode, a coalesced 3 TL $Ga_2Se_2$ crystalline film with RMS roughness as low as 0.61 nm was achieved. Unfortunately, the resulting films have undesired 3D features that are most likely caused by $Ga_2Se_3$ inclusions. Ultrasonic cleaning was used to remove most of the 3D features, after which we saw triangular features that are typical of the growth of $Ga_2Se_2$. Overall, this work describes the challenges associated with the growth of $Ga_2Se_2$ on sapphire and presents some general solutions, enabling the growth of thin $Ga_2Se_2$ films with low surface roughness.

## ACKNOWLEDGMENTS

The authors acknowledge funding from the Coherent / II-VI Foundation. The authors acknowledge the use of the Materials Growth Facility (MGF) at the University of Delaware, which is partially supported by UD-CHARM, a National Science Foundation MRSEC under Award No. DMR2011824. The co-authors acknowledge the Penn State Materials Characterization Lab for use of the PANalytical 4-Circle X'Pert3 MRD and Dr. A Richardella for helpful discussions on data interpretation.

## AUTHOR DECLARATIONS
### Conflict of Interest
The authors have no conflicts to disclose.

## DATA AVAILABILITY
The data that support the findings of this study are available at DOI: 10.5281/zenodo.7459828.

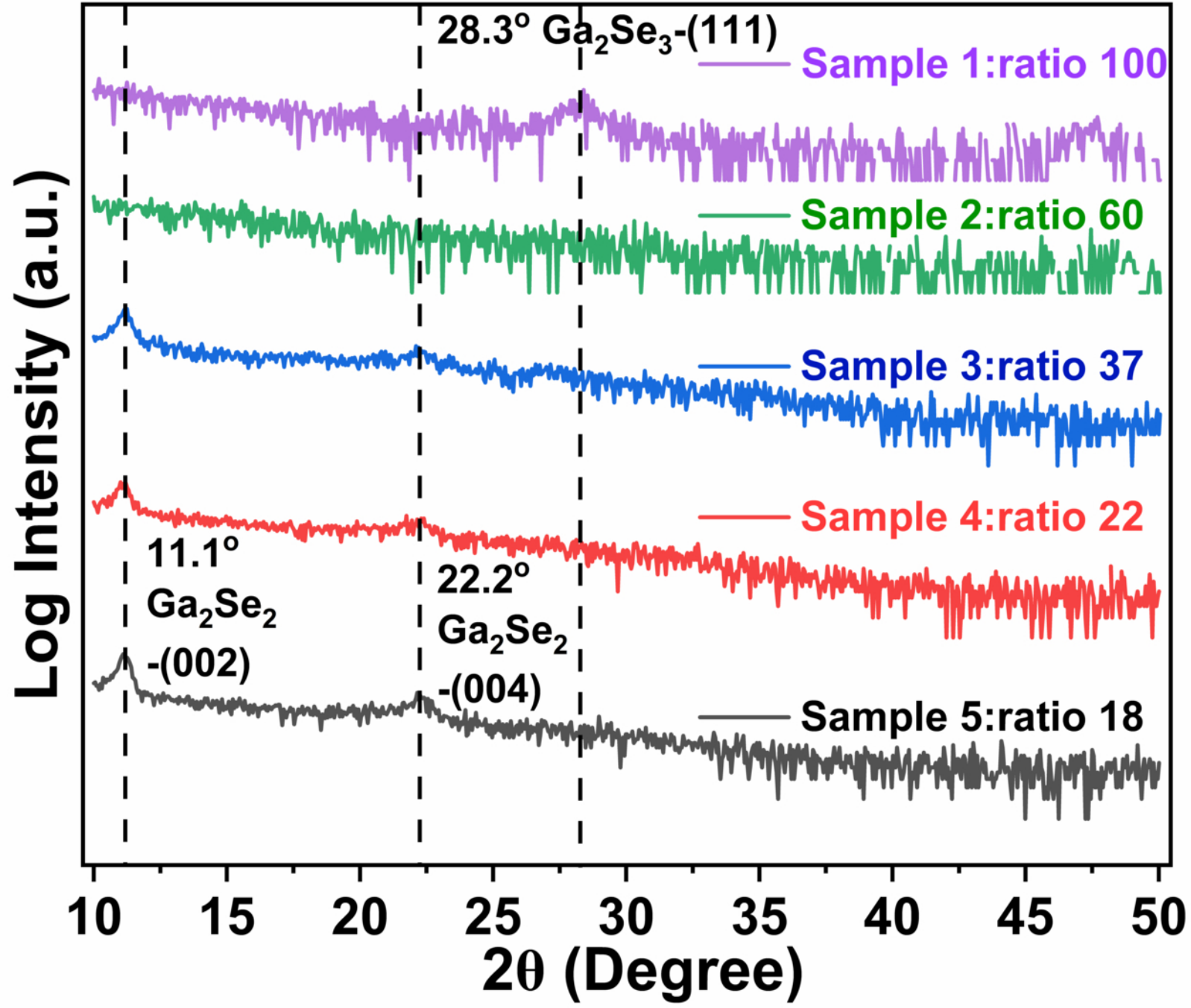

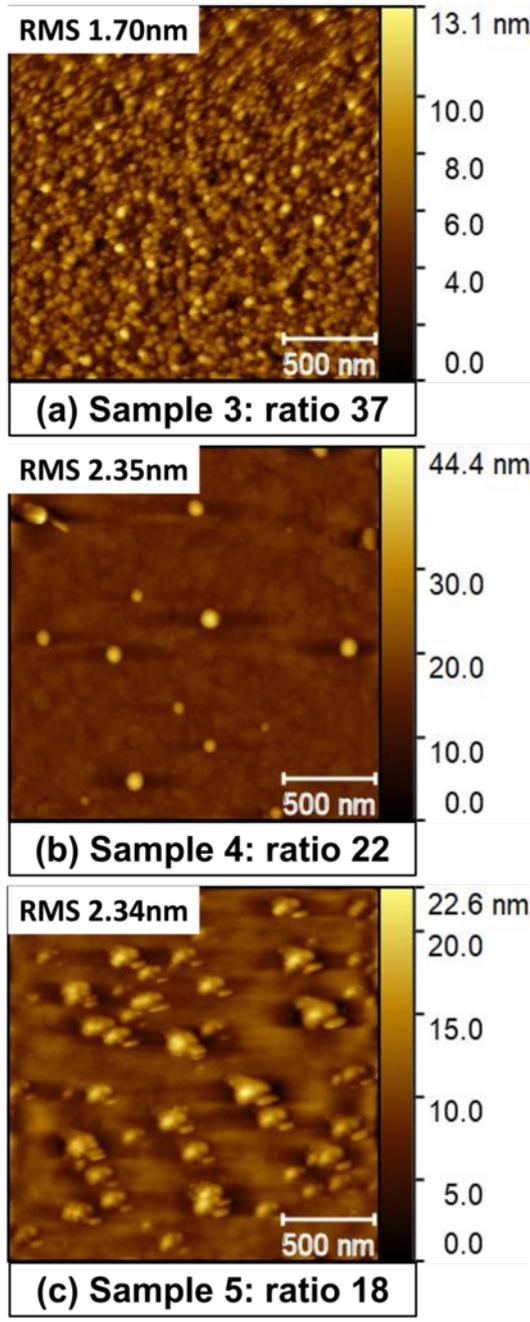

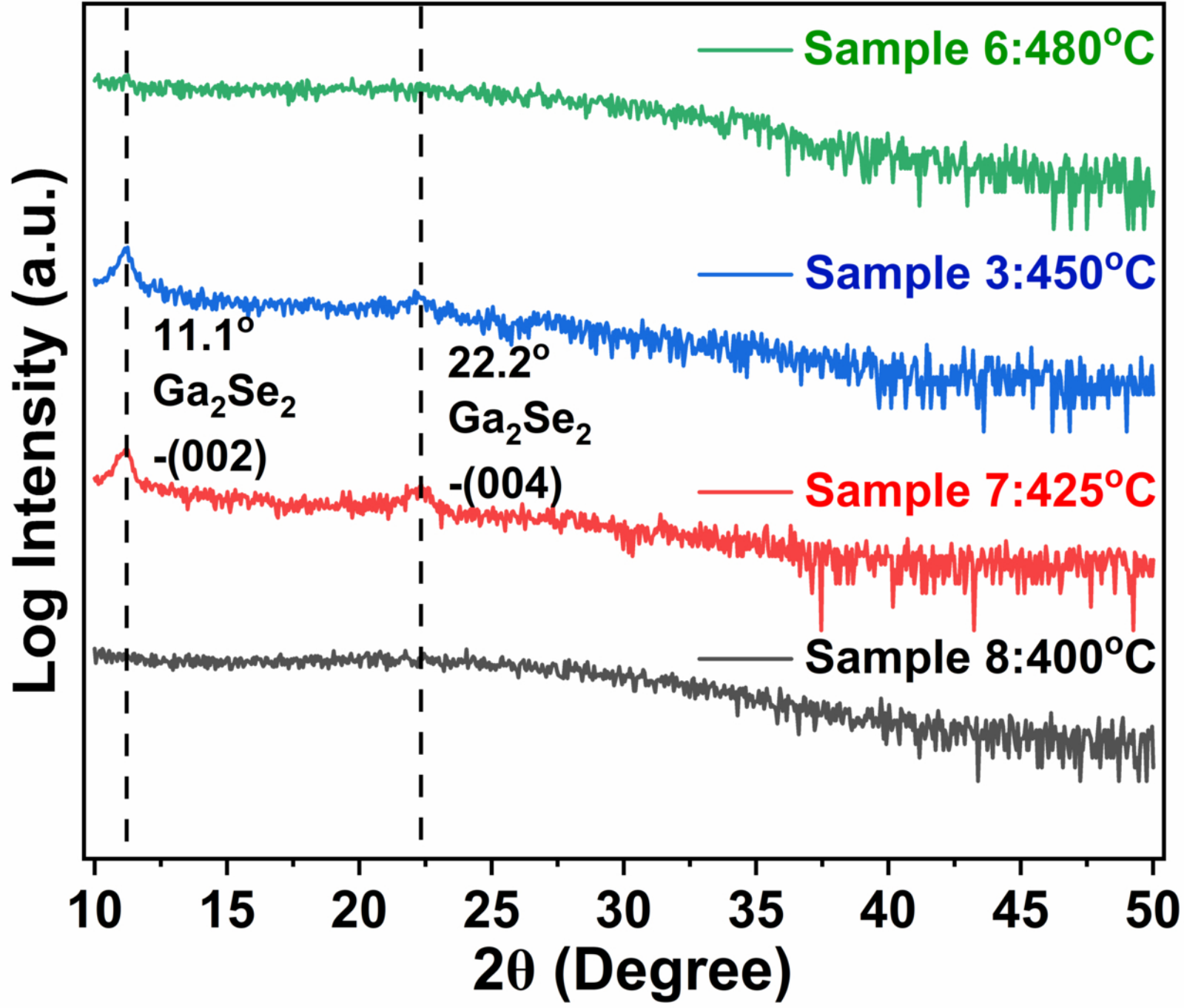

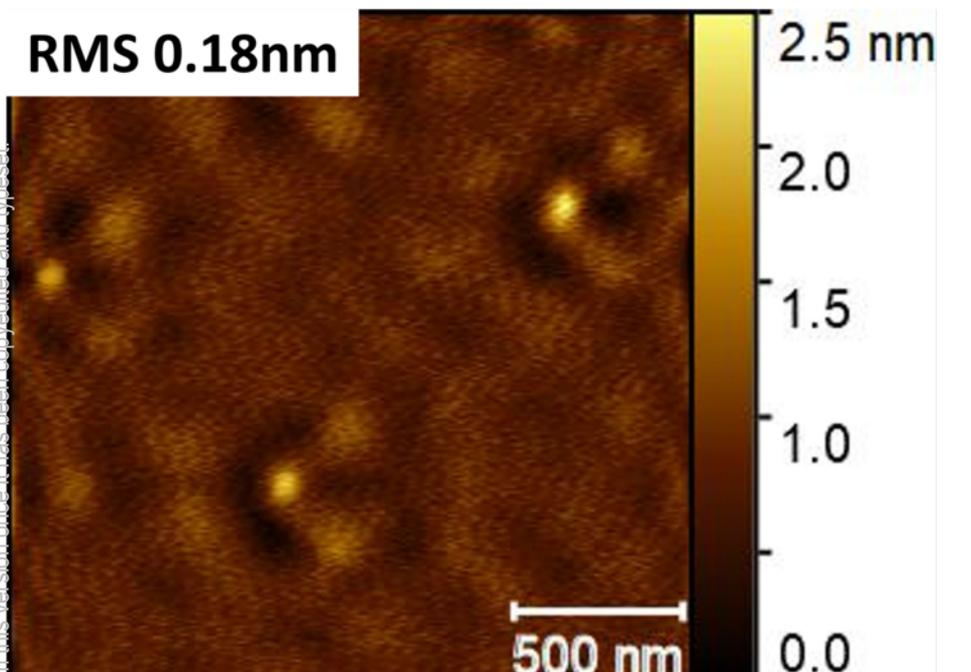
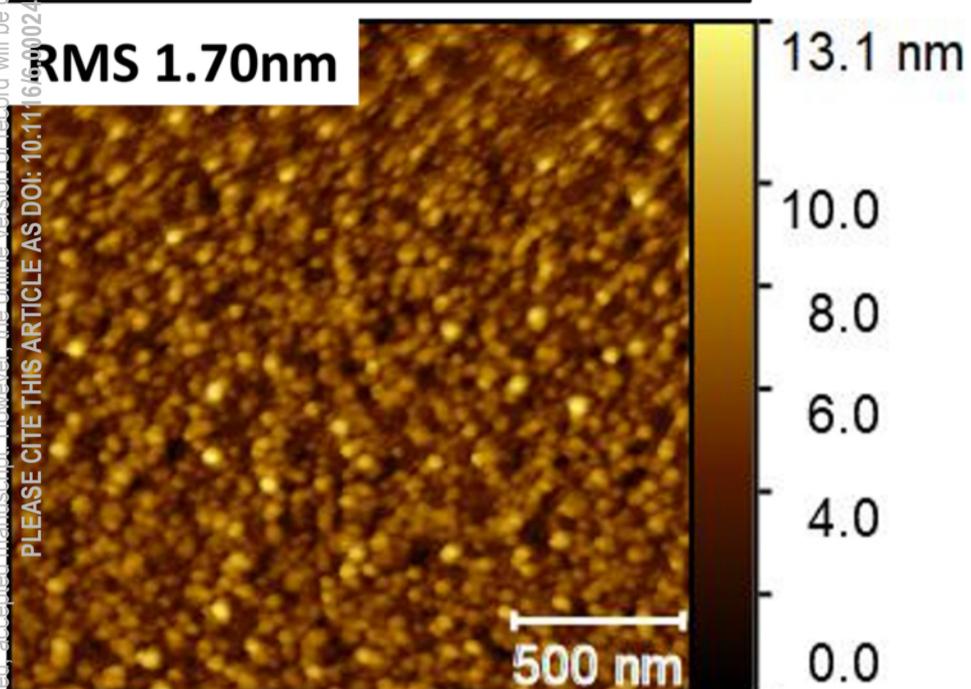
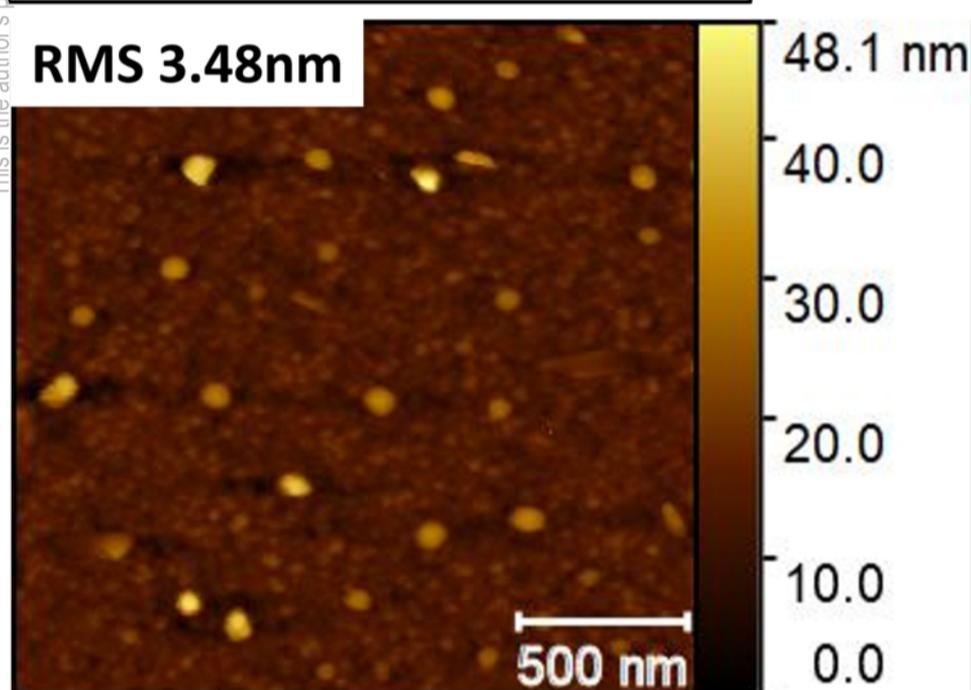
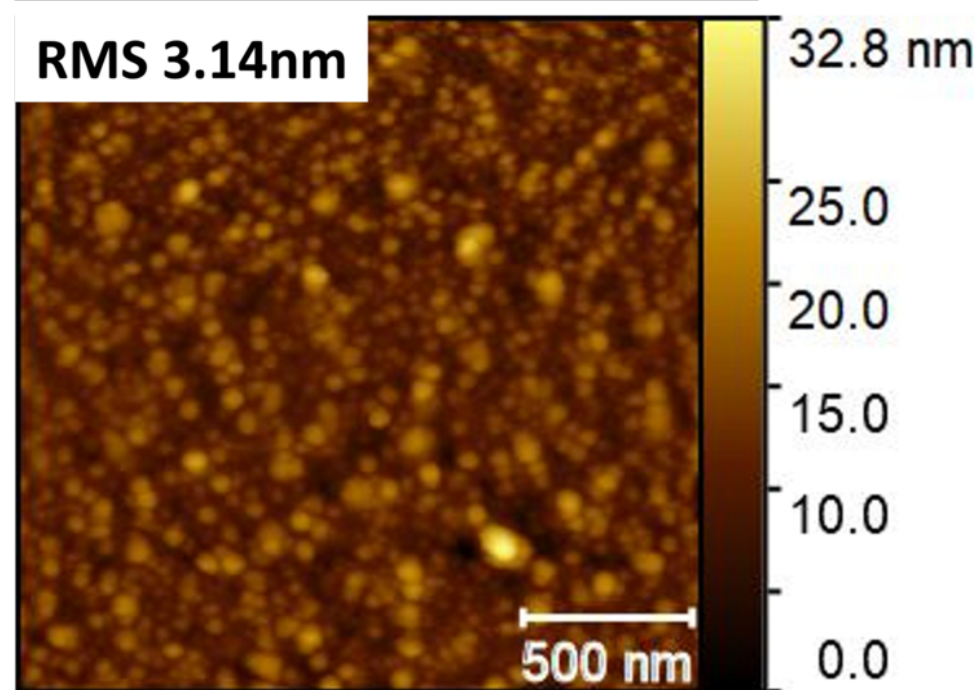
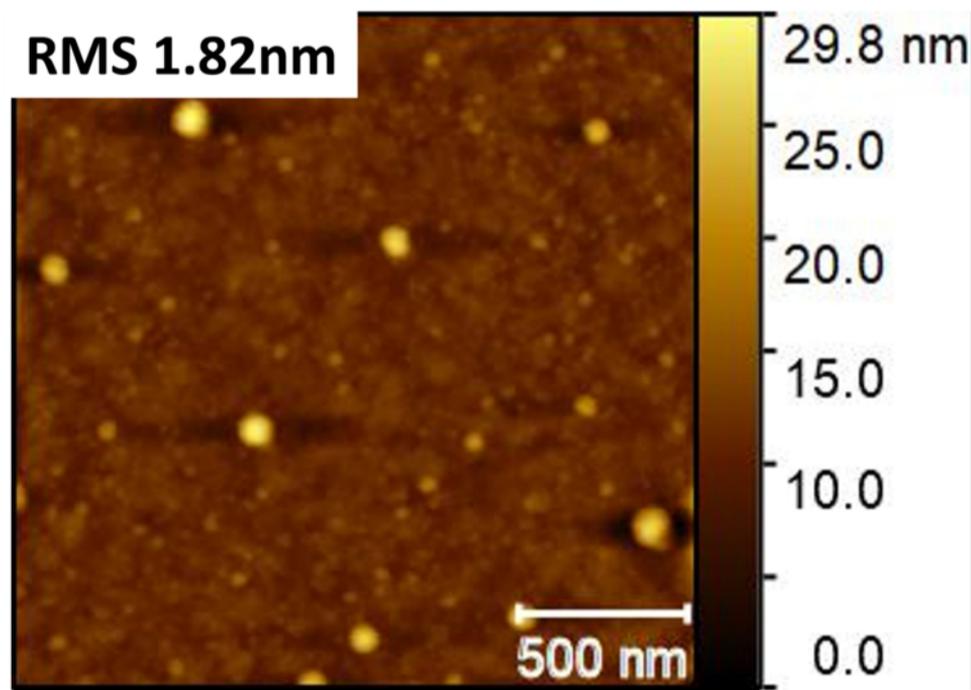

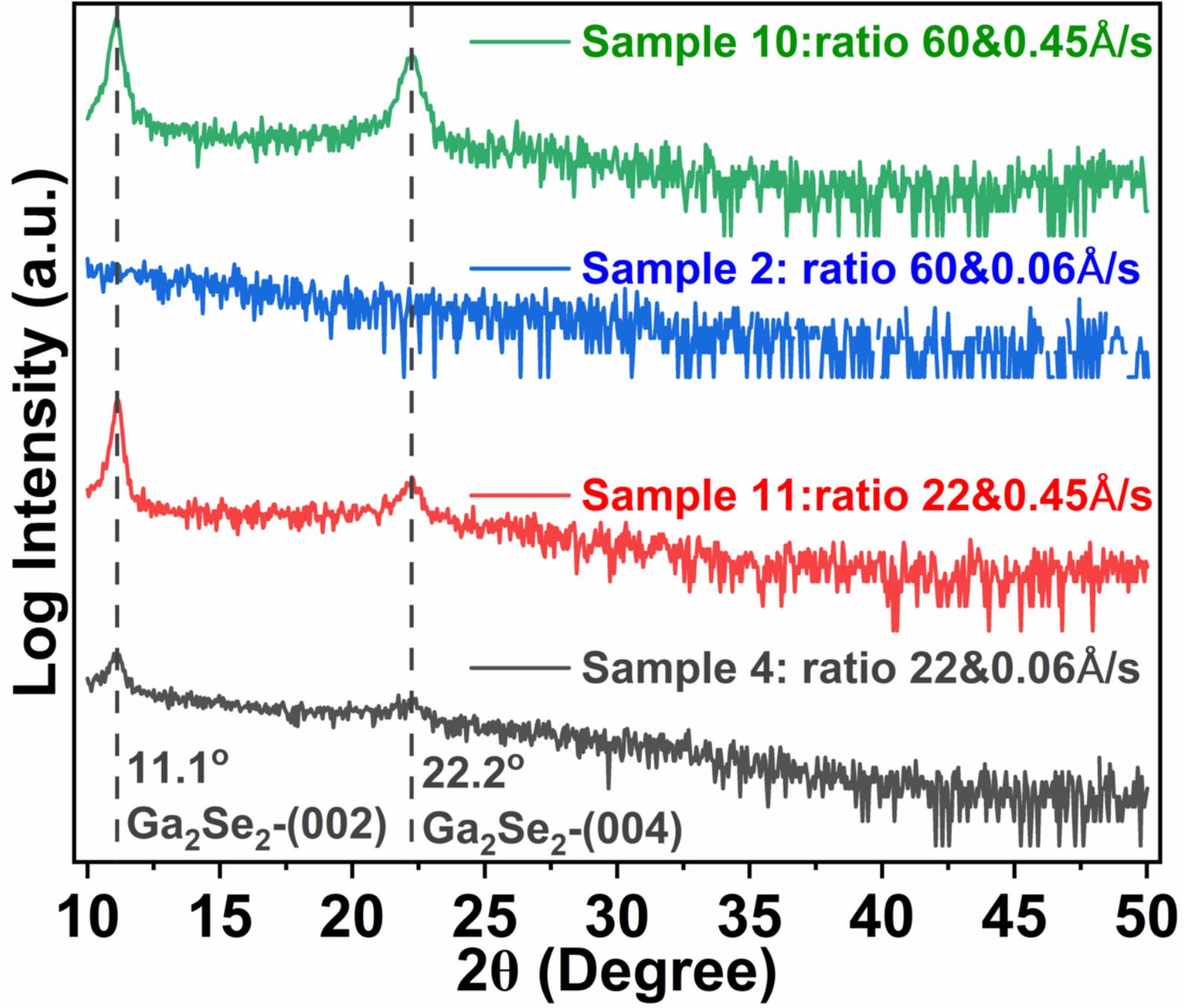

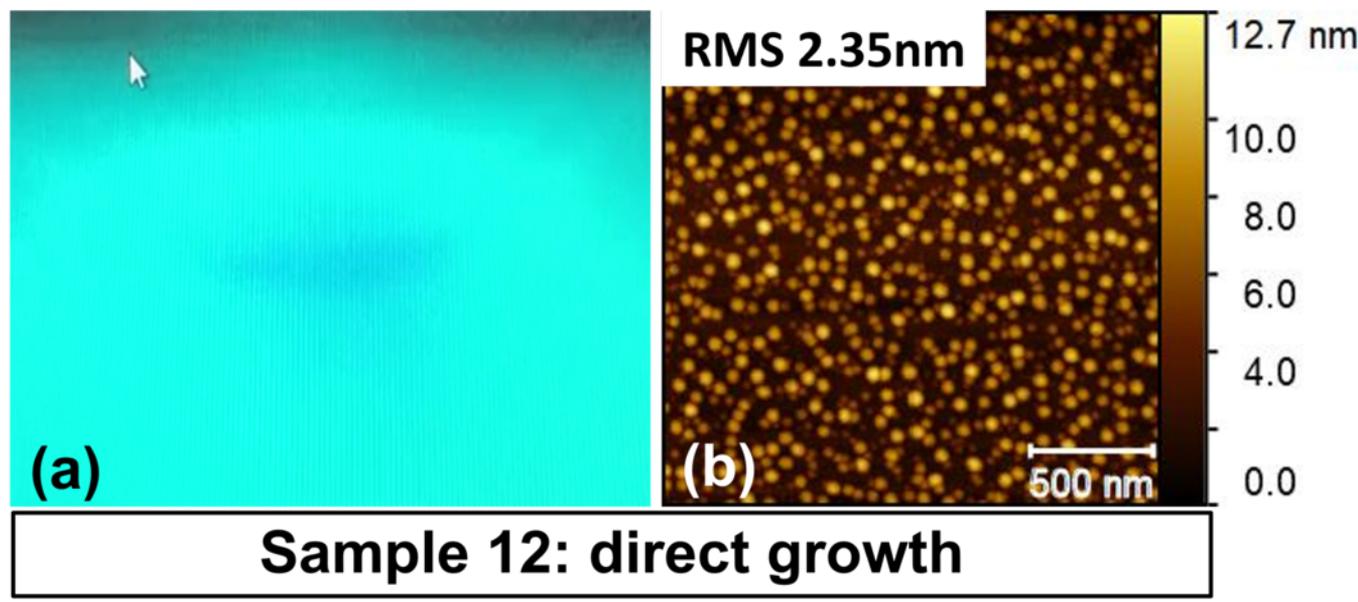
Sample 12: direct growth

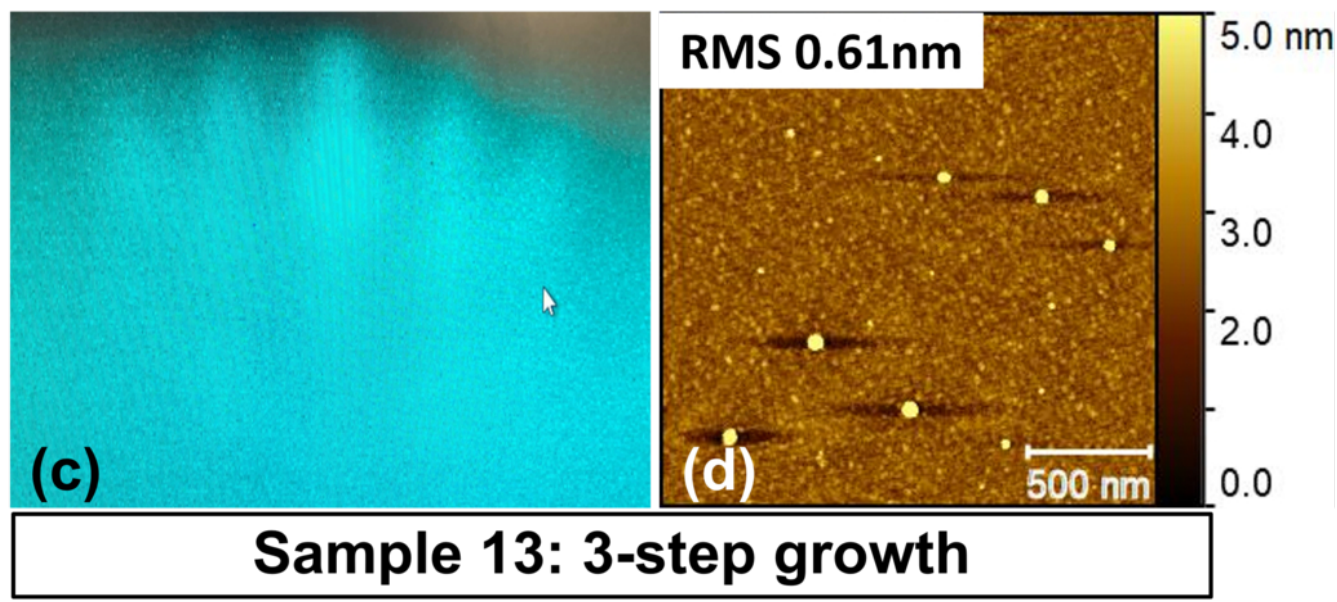
Sample 13: 3-step growth

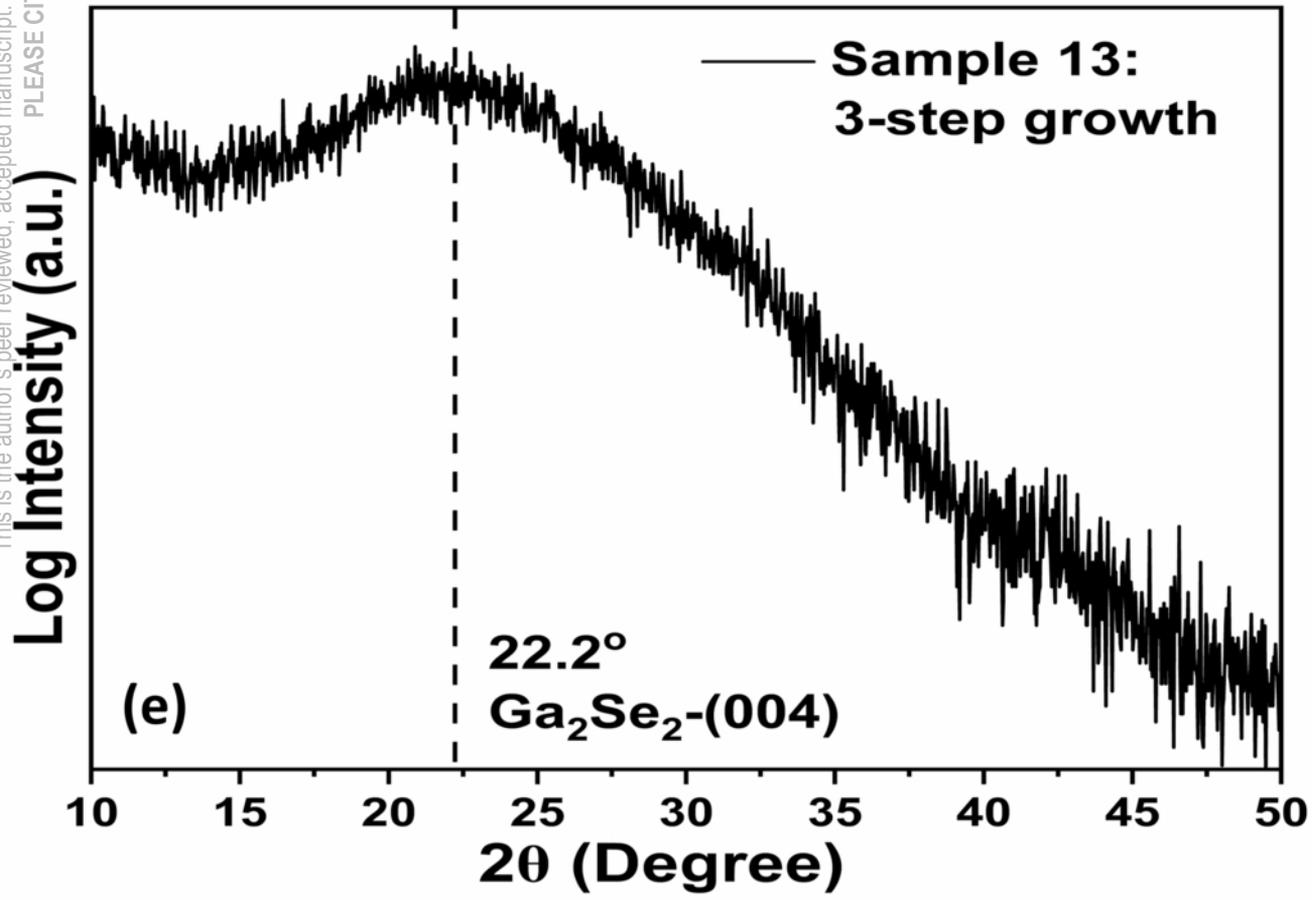

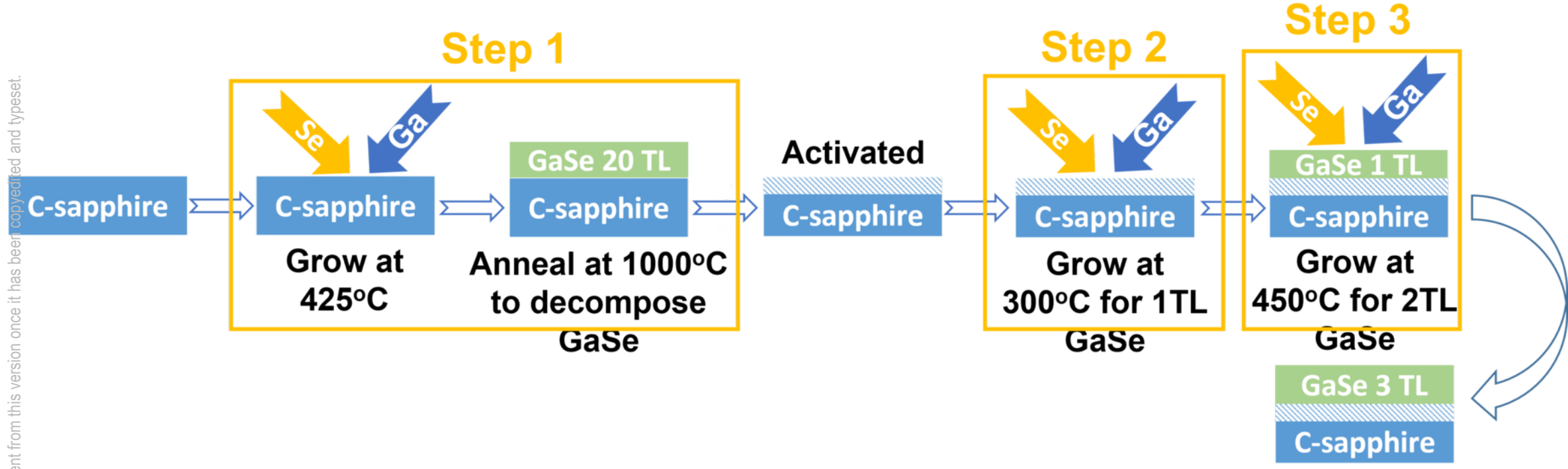

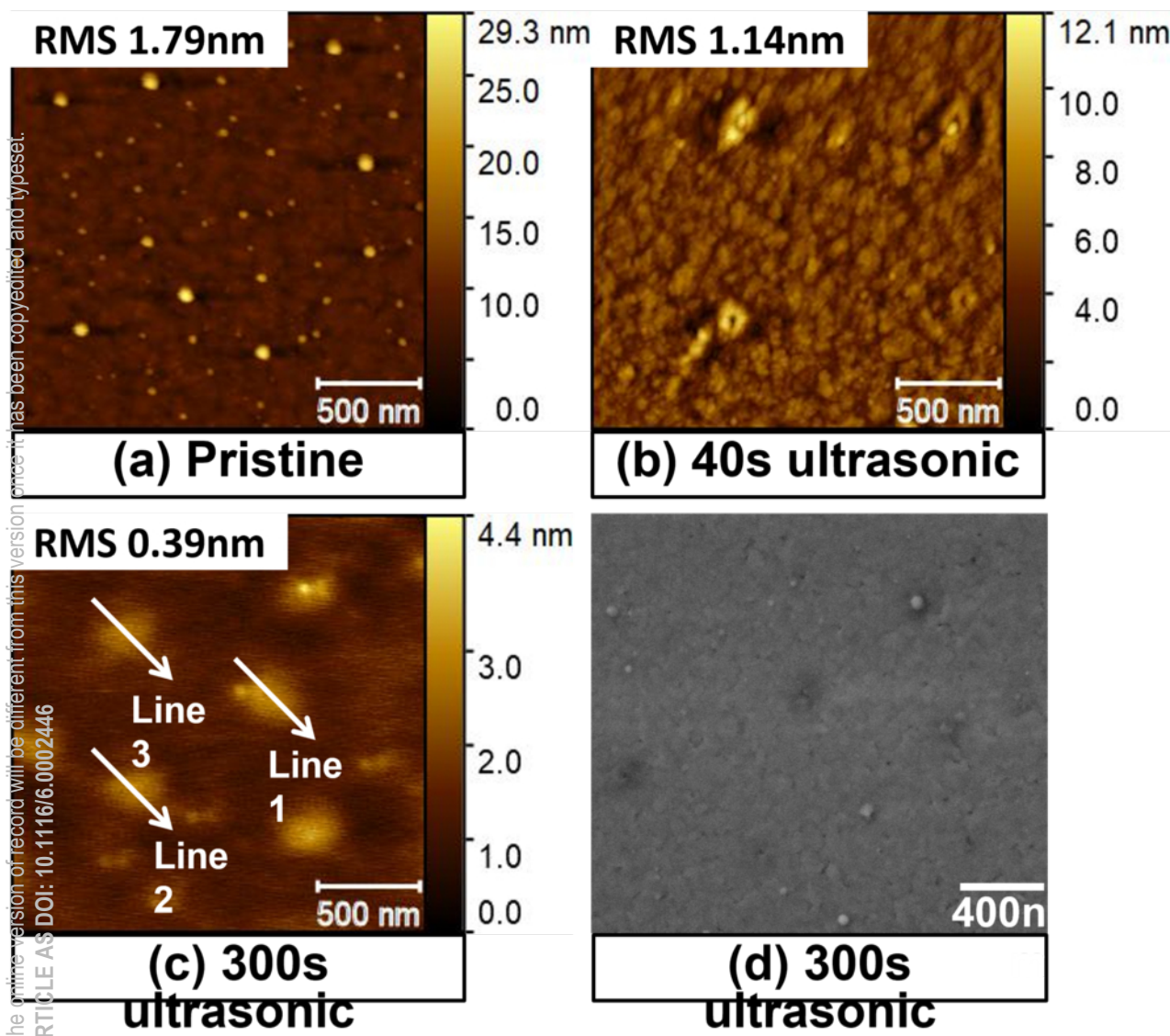

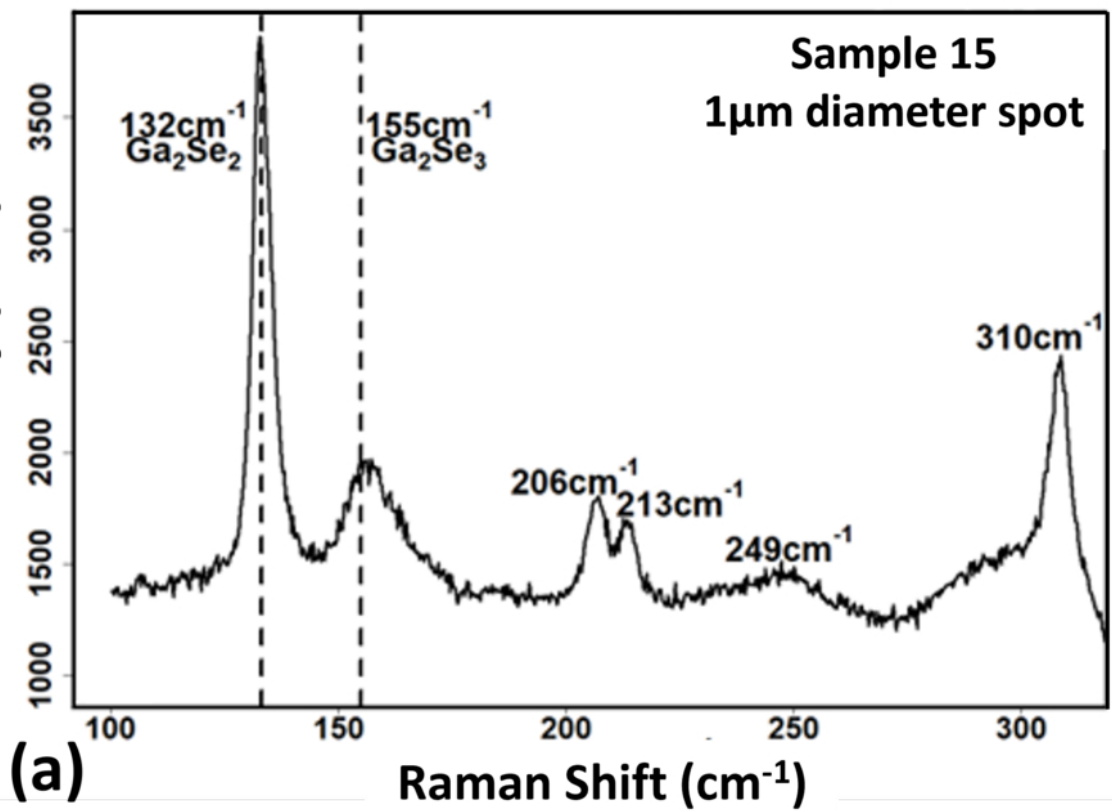
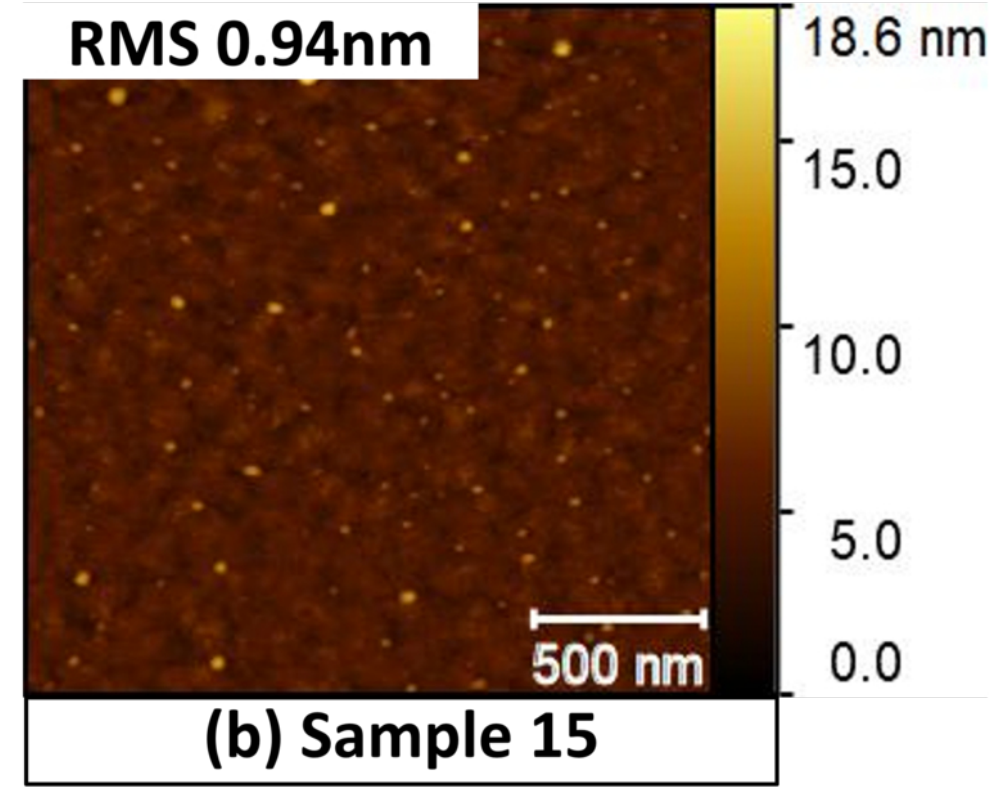
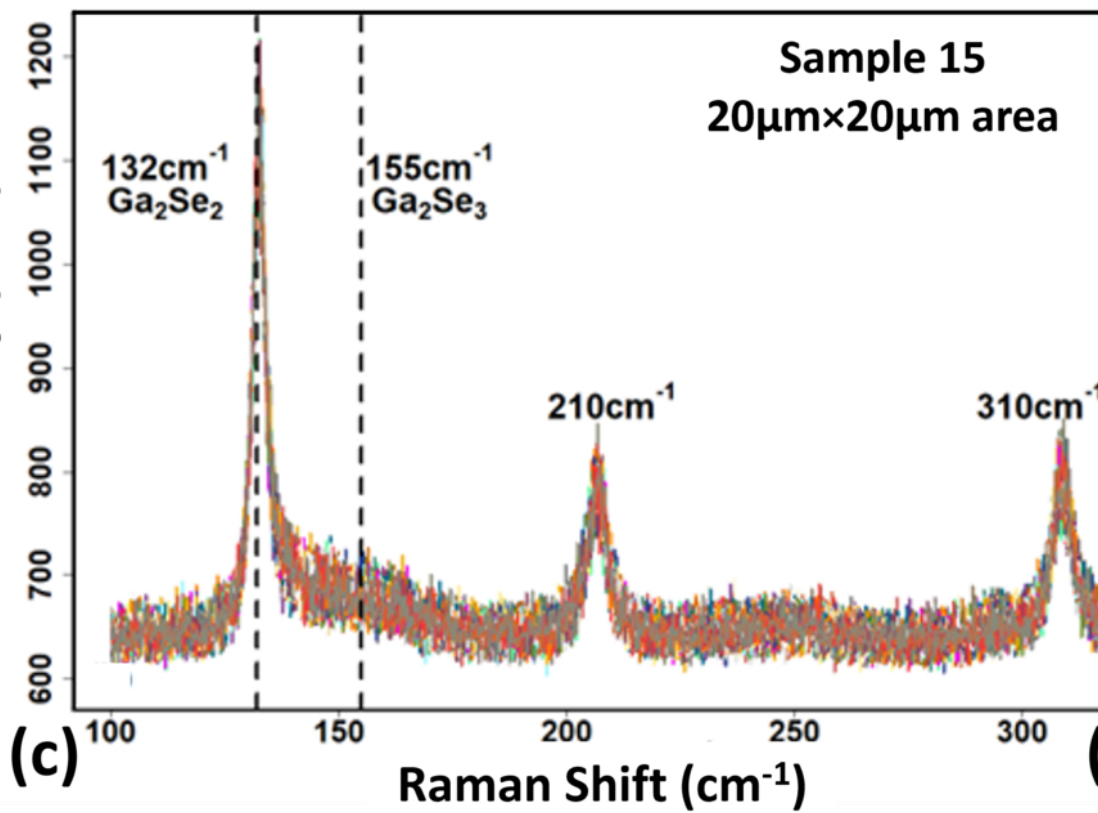
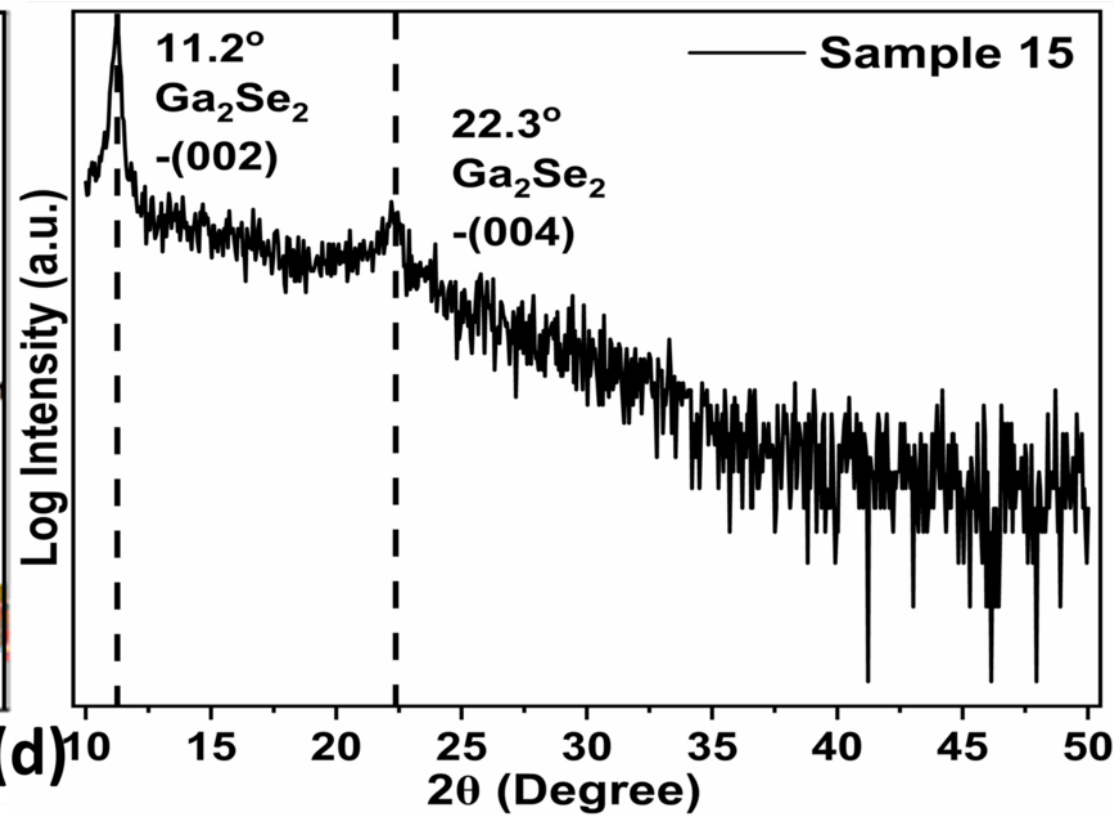